\documentstyle[spie]{article} 
\input{psfig}   

\title{Development of Prototype Pixellated PIN CdZnTe Detectors} 

\author{T. Narita\supit{1}, P. Bloser\supit{1}, J. Grindlay\supit{1}, R. Sudharsanan\supit{2}, C. Reiche\supit{2}, C. Stenstrom\supit{2} 
\skiplinehalf 
\supit{1}Harvard-Smithsonian Center for Astrophysics, 60 Garden St., Cambridge, MA 02138 
\skiplinehalf 
\supit{2}Spire Corporation, One Patriots Pk., Bedford, MA 01730
}

\authorinfo{Further author information: (Send correspondence to
T. Narita)\\T.N.: tnarita@cfa.harvard.edu\\R.S.: sudhan@spirecorp.com}
\begin{document} 
  \maketitle 

%%%%%%%%%%%%%%%%%%%%%%%%%%%%%%%%%%%%%%%%%%%%%%%%%%%%%%%%%%%%% 
\begin{abstract}
We report initial results from the design and evaluation of two
pixellated PIN Cadmium Zinc Telluride detectors and an ASIC-based
readout system.  The prototype imaging PIN detectors consist of
$4\times4$ 1.5 mm square indium anode contacts with 0.2 mm spacing and
a solid cathode plane on $10\times10$ mm CdZnTe substrates of
thickness 2 mm and 5 mm.  The detector readout system, based on low
noise preamplifier ASICs, allows for parallel readout of all channels
upon cathode trigger.  This prototype is under development for use in
future astrophysical hard X-ray imagers with 10-600 keV energy
response.  Measurements of the detector uniformity, spatial
resolution, and spectral resolution will be discussed and compared
with a similar pixellated MSM detector.  Finally, a prototype design
for a large imaging array is outlined.
\end{abstract}

%>>>> Please include a list of keywords after the abstract 


%%%%%%%%%%%%%%%%%%%%%%%%%%%%%%%%%%%%%%%%%%%%%%%%%%%%%%%%%%%%%
\section{INTRODUCTION}
\label{sect:intro}  % \label{} allows reference to this section

Much effort has been invested in recent years into fabricating
effective wide bandgap compound semiconductor detectors such as
CdZnTe.  These detectors have shown better energy resolution than
scintillators and can be made position-sensitive by adding collecting
orthogonal strips or pixels. The wide bandgap also allows these
detectors to be operated at room temperature.  However, the efficiency
and the energy resolution of the compound semiconductor detectors are
limited by the charge carrier trapping and the poor mobility-lifetime
($\mu\tau$) products for the holes.  To improve these detectors, there
is now increased effort into devising charge collection schemes that
derive the X-ray signal primarily from electron
transport\cite{barrett95,luke95}.

One such technique uses an array of anode collecting pixels with
relatively small dimensions compared to the detector thickness.  In
this so-called ``small-pixel regime'', the induced charge signal is
mostly due to the motion of the electrons in the vicinity of a
pixel, thereby reducing the dependence on the poor hole
transport\cite{barrett95}.  We have shown in our previous work with
pixellated metal-semiconductor-metal (MSM) CdZnTe detectors that a
thick (5 mm) detector with relatively large pixels (1.5 mm square)
gives good spatial and energy resolution in the 50 to 200 keV energy
band\cite{bloser97}.  One of the questions we investigate here is
whether detectors with large pixels on thick substrates are indeed
operating in the small-pixel regime.

Our work with thick CdZnTe detectors having relatively large pixels is
motivated by the need for a large area hard X-ray astronomical survey
telescope\cite{grindlay98}.  A wide-field instrument capable of imaging with
arcminute angular resolution in the 10 - 600 keV energy band is
required for Gamma Ray Burst (GRB) localization and deep surveys of
non-thermal astronomical systems.  A large area detector is needed for
imaging in the hard X-ray band since focusing optics cannot be used at
energies greater than $\sim80$ keV. Instead, one must rely on modulated
imaging techniques such as coded aperture imaging.

Our hard X-ray telescope (EXITE) group at Harvard is working
collaboratively with Spire Corporation to build and characterize
pixellated CdZnTe detectors with low leakage current.  We are
experimenting with the addition of PIN blocking contacts on CdZnTe to
reduce the bulk leakage current observed on standard MSM detectors
without blocking contacts\cite{sudhan97}.  PIN contacts have already
been used with CdTe substrate with good success\cite{}.  If PIN
contacts can improve the performance of CdZnTe, that may allow us to
build economically large detector arrays using inexpensive lower grade
CdZnTe.  In this paper, we will discuss initial results from our
program to develop pixellated PIN contact CdZnTe for future use in a
hard X-ray imaging telescope.

%%%%%%%%%%%%%%%%%%%%%%%%%%%%%%%%%%%%%%%%%%%%%%%%%%%%%%%%%%%%%
\section{CZT P-I-N DETECTOR DEVELOPMENT} 
\label{sect:CZT_PIN}  % \label{} allows reference to this section

%%-----------------------------------------------------------
\subsection{Detector and Readout Electronics} 

\begin{figure}
\begin{center}
\begin{tabular}{c}
\psfig{figure=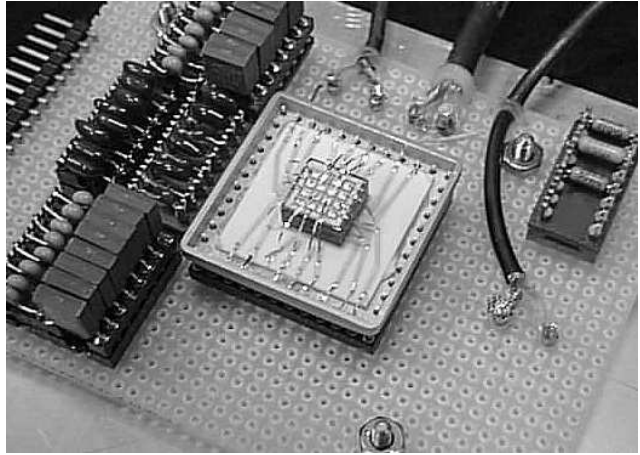,height=6cm} 
\end{tabular}
\end{center}
\caption[pict] 
{ \label{fig:pict}	  
The 5 mm PIN CdZnTe detector.  The wirebonds
are connected to the anode pixels.  The detector is illuminated from
underneath the chip carrier (cathode side). } 
\end{figure}

Our detectors are spectroscopic grade High Pressure Bridgeman CdZnTe
with 10\% Zn from eV Products. The substrate sizes are
$10\times10\times2$ mm and $10\times10\times5$ mm.  $1 \mu$m layers of
CdS (n-type) and ZnTe (p-type) are deposited on to the CdZnTe surface
by thermal evaporation to form a diode-like blocking contact
(PIN)\cite{sudhan97}.  A mask is used on the CdS side to form 16
$1.5\times1.5$ mm square pixels with 0.2 mm gaps between pixels.  A 1 mm wide guard
ring is also placed around the array of pixels with a 0.2 mm gap
separating the array from the guard ring.  A layer of indium is
finally deposited over the CdS to form the metal contact.
Figure~\ref{fig:pict} shows the pixel side of the detector and the
discrete components of the readout electronics. 

The CdZnTe detector is attached to a ceramic substrate and mounted in
a chip carrier.  A window in the metal chip carrier allows X-rays to
impinge on the cathode surface through the ceramic substrate, which
imposes a low energy cutoff at $\sim20$ keV.  Wirebonds are used to
connect the pixels to the pins of the chip carrier.  The cathode side
of the detector is connected to an eV Products 550 preamplifier and
biased to a large negative voltage.  The output of the preamp is
shaped by a NIM module fast shaping amplifier and used as an event
trigger.  The anode pixels are biased to ground via 100 M$\Omega$
resistor and AC coupled to two 8 channel charge sensitive
preamplifer--shaping amplifier Application Specific Integrated
Circuits (VA-1 model ASIC) manufactured by Integrated Device
Electronics (IDE).  A shaping time of $1\mu$s is used and the ASIC
response is measured to be linear over the energy range of interest.
The ASIC outputs are amplified and simultaneously held when triggered
by the fast shaped cathode pulse.  The 16 held signals are converted
by a PC/104 single-board-computer and a daughter board ADC card, and
written out via parallel port interface to a PC on an event-by-event
basis.  To determine the noise in the system due to the ASIC, stray
capacitance, and leakage current, test pulses ($\sim1$ mV) can be
injected through 2 pF capacitors into the ASIC in parallel with the
detector output.  The gains of the post-ASIC amplifiers are not
identical, but we have calibrated each channel using known X-ray
lines.

%%-----------------------------------------------------------
\subsection{Leakage Current} 
\label{sect:leakage}  % \label{} allows reference to this section

\begin{figure}
\begin{center}
\begin{tabular}{cc}
\psfig{figure=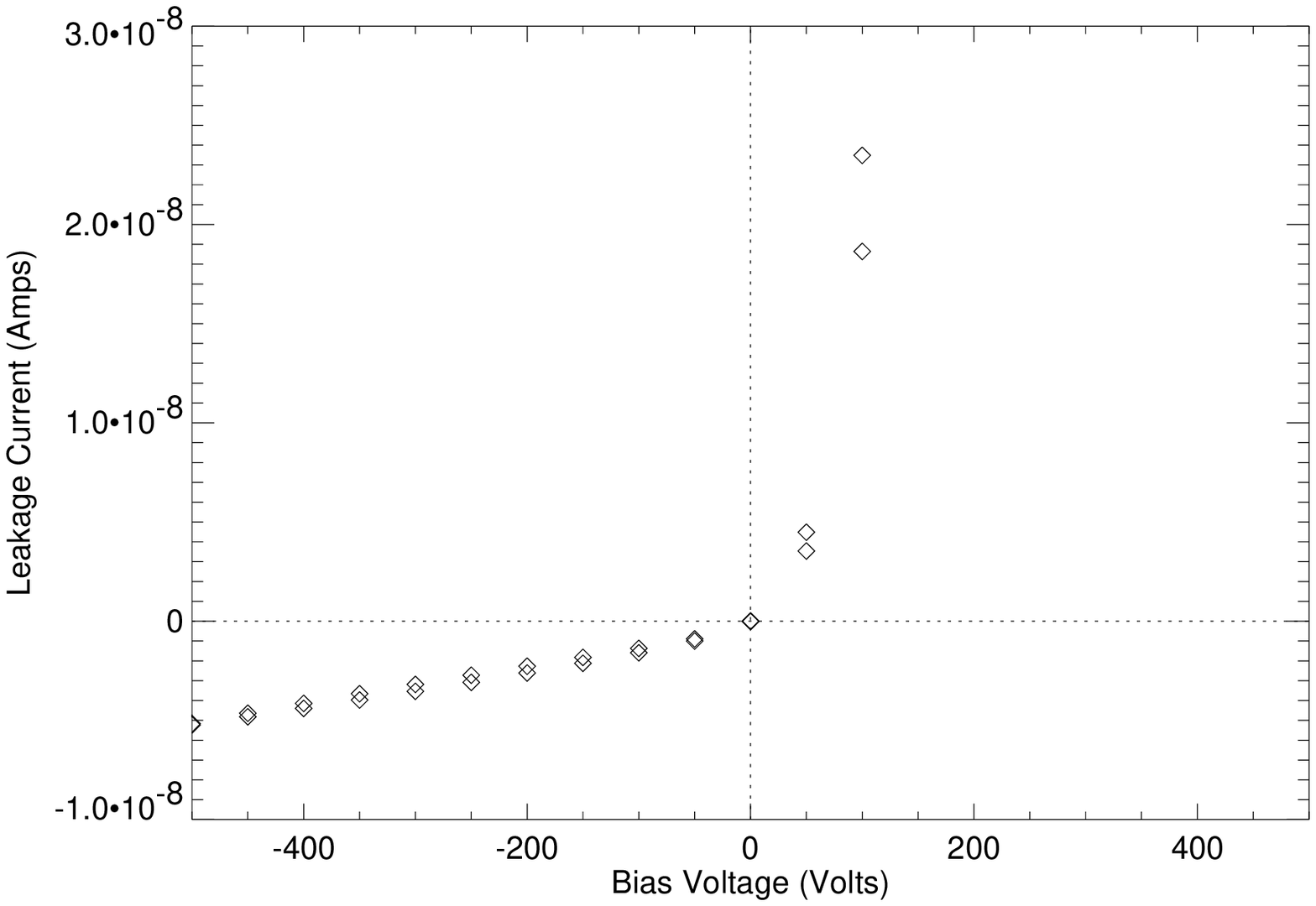,height=5cm} 
\psfig{figure=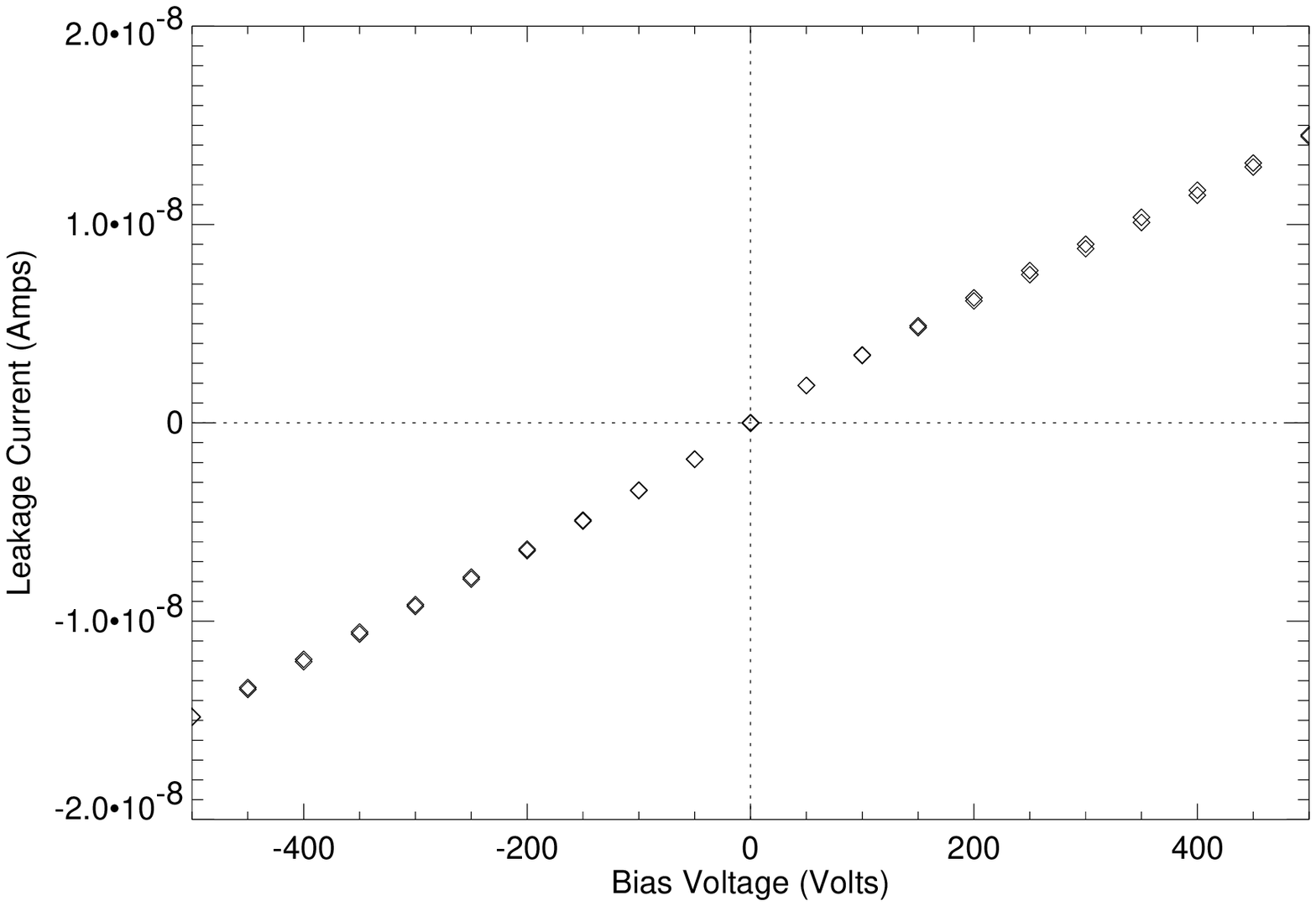,height=5cm} 
\end{tabular}
\end{center}
\caption[leakage] 
{ \label{fig:leakage}	  
Comparison of the leakage current between PIN and MSM detectors (each
5 mm thick).  At -500 volts bias, the PIN detector shows 5 nA of
leakage current while the MSM detector shows 15 nA.} 
\end{figure}

Since the bulk leakage current contributes to the signal noise and
limits the operating detector bias, we compared the leakage currents
between PIN and MSM detectors.  We used a Keithley 237 high voltage
source to bias each pixel of the PIN detector in 50 volt steps from
+500 to -500 volts.  The leakage current measurements were taken in
the dark and at room temperature.  The typical leakage current across
the 2 mm thick CdZnTe was 10 nA at -500 volts and the current across
the 5 mm thick detector was 5 nA at -500 volts.  The leakage current
from a similarly pixellated $10\times10\times5$ mm MSM
detector\cite{bloser97} at equal bias was 15 nA
(Fig.~\ref{fig:leakage}).  The PIN detector clearly exhibits the
standard diode current-voltage relation, confirming the effectiveness
of blocking contacts on reducing the bulk leakage current.  The bulk
resistivity of the PIN detector is estimated to be $2\times10^{11}$
ohms-cm, while the MSM detector without the blocking contact has a
resistivity of only $6.7\times10^{10}$ ohms-cm.  Additional
experiments with various surface passivations are planned to further
reduce the leakage current to few nanoamps at comparable bias voltage.

The surface leakage current contribution was also measured by placing
a small difference in potential between an edge pixel and the normally
grounded guard ring (separated from the pixel by the 0.2 mm gap).  The
typical surface current was 0.8 nA at 5 volts for the PIN detector and
1 nA at 1 volt for the MSM detector.  Thus the surface leakage current
in the PIN detector is significantly reduced, which has important
advantages for guard ring designs, as discussed below (Sec.~\ref{sect:guardbias}).

The charge transport property of a semiconductor is one of the
important parameters in determining the detector performance.  Large
values for the charge carrier mobility ($\mu$) and the mean drift time
($\tau$) are sought for good charge collection.  We measured the
electron $\mu\tau$ product for the PIN detectors and the MSM detector
by irradiating the cathode contact with a $^{241}$Am source (60 keV)
and recording the peak pulse height amplitude value for pulses shaped
with $1\mu$s time constants at various bias voltage values.  By
fitting the result with the Hecht relation\cite{hecht32}, we find that
typical $\mu\tau$ for the MSM detector is $2.8\times10^{-3} cm^{2}
V^{-1}$, whereas the PIN detector $\mu\tau$ are $5.1\times10^{-3}
cm^{2} V^{-1}$ and $2.0\times10^{-3} cm^{2} V^{-1}$ for the 2 mm and
the 5 mm respectively.

%%-----------------------------------------------------------
\subsection{Energy Resolution} 

\begin{figure}
\begin{center}
\begin{tabular}{cc}
\psfig{figure=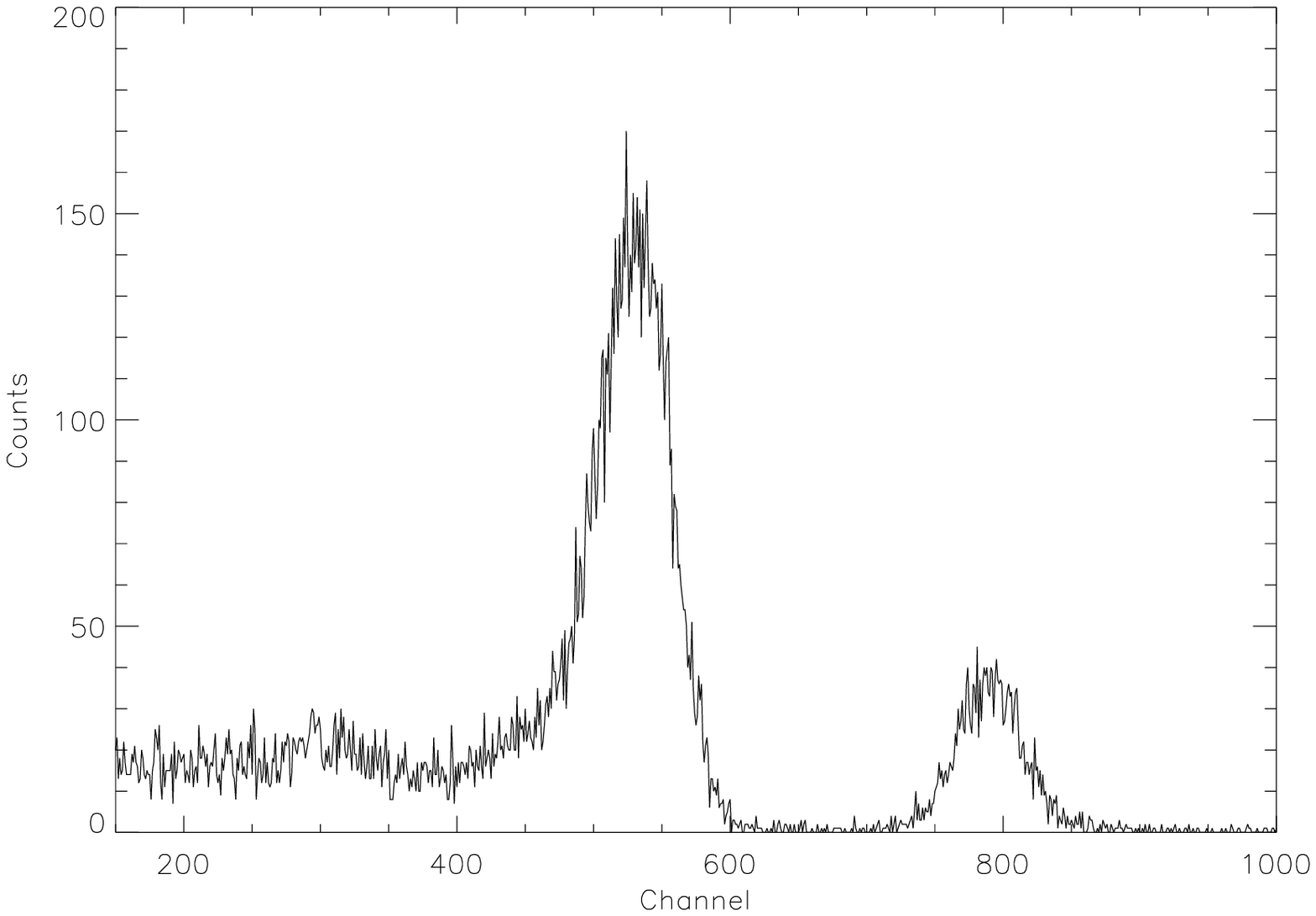,height=5cm} 
\psfig{figure=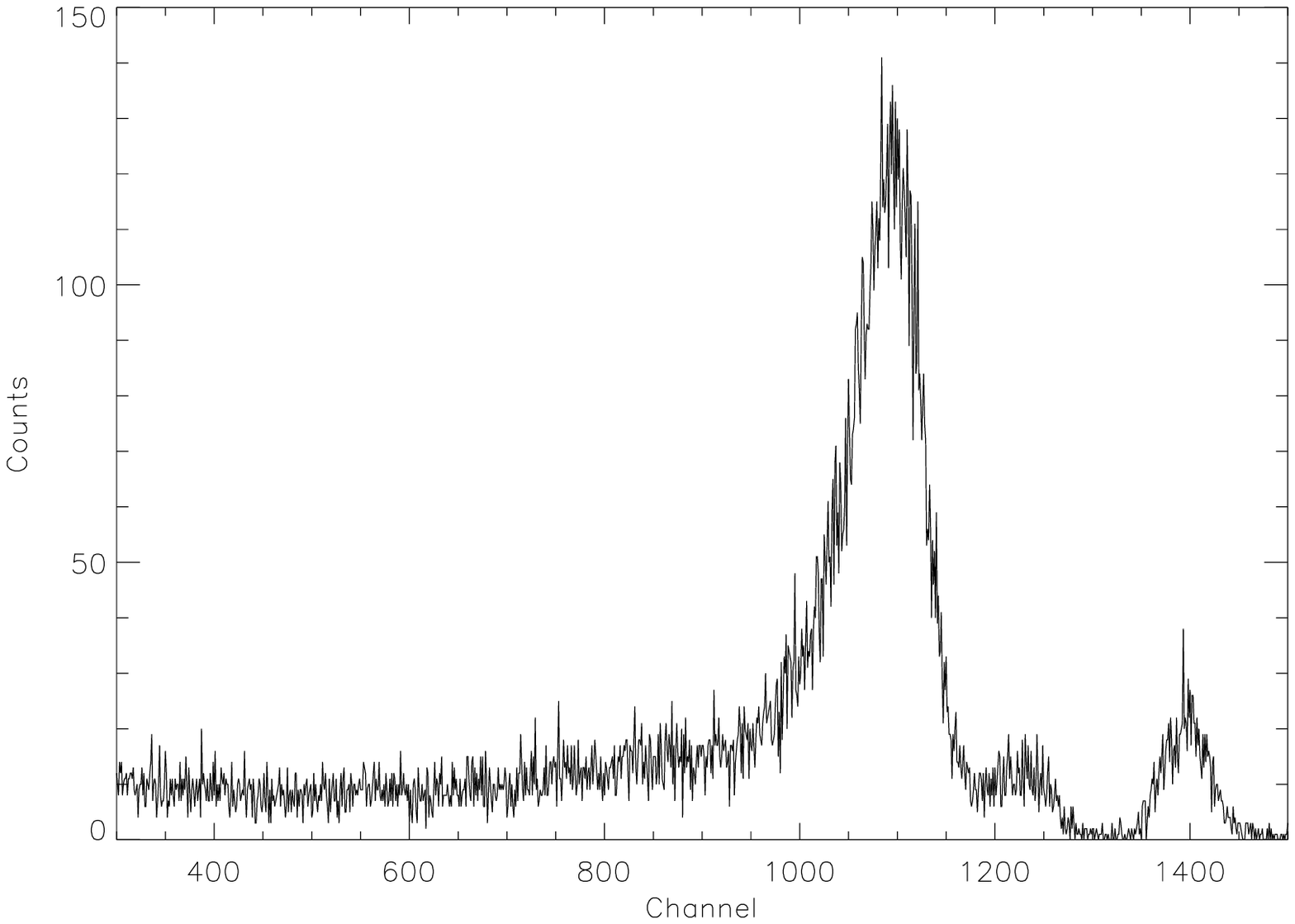,height=5cm} 
\end{tabular}
\end{center}
\caption[energy] 
{ \label{fig:energy}	  
Typical spectra from the 5 mm PIN detector biased at -500
volts. $^{241}$Am (60 keV) and $^{57}$Co (122 and 136 keV) spectra are
shown in the left and right spectra, respectively, together with
their pulser peaks (rightmost peak in each spectrum) for comparison.} 
\end{figure}

A primary concern when using a CdZnTe substrate is the uniformity of
the energy resolution across the detector.  Small structural defects
in CdZnTe can trap charge carriers and thereby degrade the local
energy resolution.  We uniformly illuminated both the 2 mm and the 5
mm detectors with $^{241}$Am and $^{57}$Co at -500 volt bias to
determine the energy resolution of all the pixels, and also measure
any depth-dependent effects.  The energy resolution is defined as the
ratio of the FWHM width of the Gaussian photopeak to the peak energy.
Test pulses were simultaneously injected to measure the electronic
noise contribution to the photopeak.  The typical energy resolution at
60 keV was $\sim7-10$\% for the 2 mm detector and $\sim9-11$\% for the
5 mm detector.  As seen from the pulser, most of the spread in the
resolution is due to electronic noise.  The primary noise contribution
is the stray capacitance from the discrete components on our detector
interface card, as also measured from the MSM detector\cite{bloser97}.
By subtracting the test pulser width in quadrature from the photopeak
width, we estimate the intrinsic detector resolution at 60 keV to be
$\sim3.2-4.8$\% for the 2 mm detector and $\sim4.5-6.5$\% for the 5 mm
detector.  At 122 keV, we estimate the intrinsic detector resolution,
following the same process of test pulser subtraction, to be
$\sim2.5-4.0$\% for the 2 mm detector and $\sim4.0-5.2$\% for the 5 mm
detector.  Figure~\ref{fig:energy} shows the typical spectrum obtained
from one pixel for both the $^{241}$Am and $^{57}$Co sources.  At the
same -500 volt detector bias, the MSM detector gave comparable energy
resolution of $\sim3.8-5.5$\% at 60 keV.

The detector energy resolution is found to be fairly uniform for
either of the detectors at two energies.  No significant
depth-dependent variation in the energy resolution was observed.
Although the escape peaks at $\sim30$ keV are visible, the lower
energy Am and Np lines below $\sim20$ keV could not be measured due to
absorption by the ceramic substrate.  The 5 mm PIN detector's energy
resolution was the worst of the three detectors we tested, but this
could be attributed to the poorer quality of the CdZnTe used, as seen
by the smaller electron $\mu\tau$ (Sec.~\ref{sect:leakage}).  At
higher bias (-700 volts), the 60 keV energy resolution of the 5 mm PIN
detector improved to $\sim3.5-5.0$\%, or comparable to the 5 mm MSM
detector at -500 volt bias.  The improvement in the energy resolution
of the PIN detector at larger bias may imply that the detector is not
fully depleted at 500 volts.

%%-----------------------------------------------------------
\subsection{Photopeak Efficiency} 

\begin{figure}
\begin{center}
\begin{tabular}{cc}
\psfig{figure=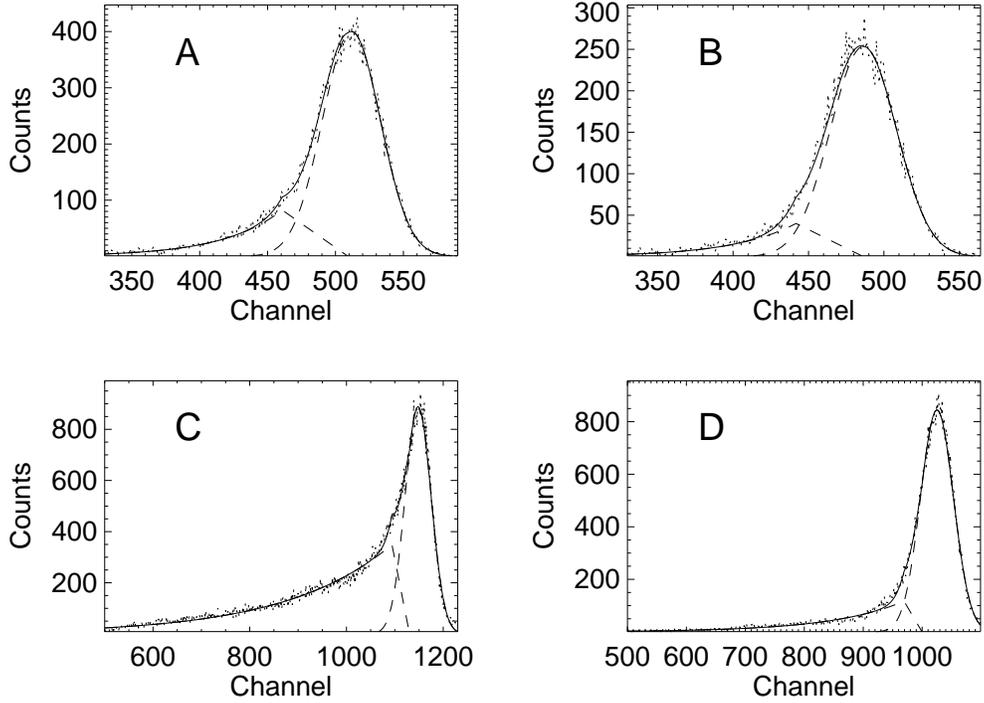,height=10cm} 
\end{tabular}
\end{center}
\caption[sm_pix] 
{ \label{fig:sm_pix}	  
Photopeak efficiency fits for the 2 mm and the 5 mm detectors at two
collimated beam energies. A, $^{241}$Am on 2 mm CdZnTe.  B,
$^{241}$Am on 5 mm CdZnTe.  C, $^{57}$Co on 2 mm CdZnTe.  D,
$^{57}$Co on 5 mm CdZnTe. The best-fit Gaussian (photopeak) and
exponential tail components are shown as dashed curves which sum to
the solid curve.}
\end{figure}

The primary benefit from operating a pixellated detector in the
small-pixel regime is the improved photopeak efficiency.  We used
$^{241}$Am and $^{57}$Co beams, collimated to $\sim0.8$ mm diameter, centered on the
pixels of the 2 mm and the 5 mm PIN detectors, biased to -500 volts,
to compare the amount of low energy tailing in the photopeaks.  By
using a collimated X-ray beam, any contribution to the photopeak tail
due to incomplete charge collection from X-ray interactions in the
interpixel region can be neglected.  Each spectrum is fit using a
$\chi^{2}$ minimizing Gaussian and an exponential tail.  We define the
photopeak efficiency as the ratio of the counts in the Gaussian
photopeak to the total counts in the Gaussian plus the exponential
tail.  Figure~\ref{fig:sm_pix} shows the spectra for the two detectors
at two X-ray energies and their respective fits.  The $^{241}$Am
photopeak efficiencies are $79.0\pm2.8\%$ and $88.9\pm4.2\%$
($1\sigma$) for the 2 mm and the 5 mm detectors, while the respective
$^{57}$Co photopeak efficiencies are $42.5\pm0.9\%$ and $77.5\pm2.5\%$.

The difference in the photopeak efficiencies between the 2 mm and 5 mm
detectors illuminated with $^{241}$Am is small.  This is expected since the
average depth of interaction by a 60 keV photon in CdZnTe is $\sim0.1$
mm.  At this small penetration depth, the small pixel advantage of
single-charge collection is minimal.  At 120 keV, the X-rays
penetrate farther ($\sim1$ mm) into the CdZnTe.  The distribution of
interactions occurring deeper in the weighting potential, and some
contribution from electron trapping, will likely produce a larger
photopeak tail compared to the $^{241}$Am X-rays.  It is clear
however, when comparing the $^{57}$Co spectrum on the 2 mm detector to
the 5 mm detector, that our 5 mm detector is indeed operating in the
small-pixel regime and maintaining a reasonable photopeak efficiency.

%%-----------------------------------------------------------
\subsection{Position Resolution}

The imaging quality of a pixellated detector is limited by charge
spreading due to ionizing photoelectron, Compton scattering of the
primary photon, and diffusion of electron-hole clouds under bias
potential between the collecting electrodes.  We used collimated and
full-flood beams to measure the point-spread-function (PSF) of our
relatively large pixel detector and to determine the best method of
recovering our signal.

\begin{figure}
\begin{center}
\begin{tabular}{c}
\psfig{figure=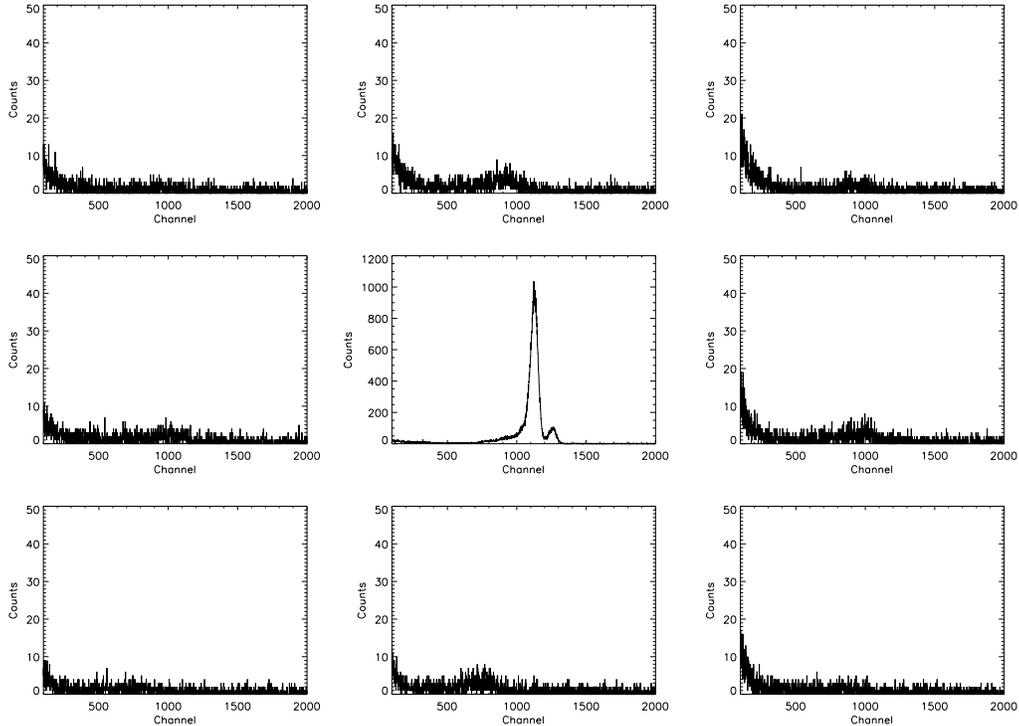,height=10cm} 
\end{tabular}
\end{center}
\caption[col_pos] 
{ \label{fig:col_pos}	  
Collimated $^{57}$Co spectrum and its nearest neighbors.  Compton
scattered photons from the 122 keV and 136 keV photons produce small
photopeaks in the neighboring pixels.
There was some variation in the gains between channels. } 
\end{figure}

\begin{figure}
\begin{center}
\begin{tabular}{cc}
\psfig{figure=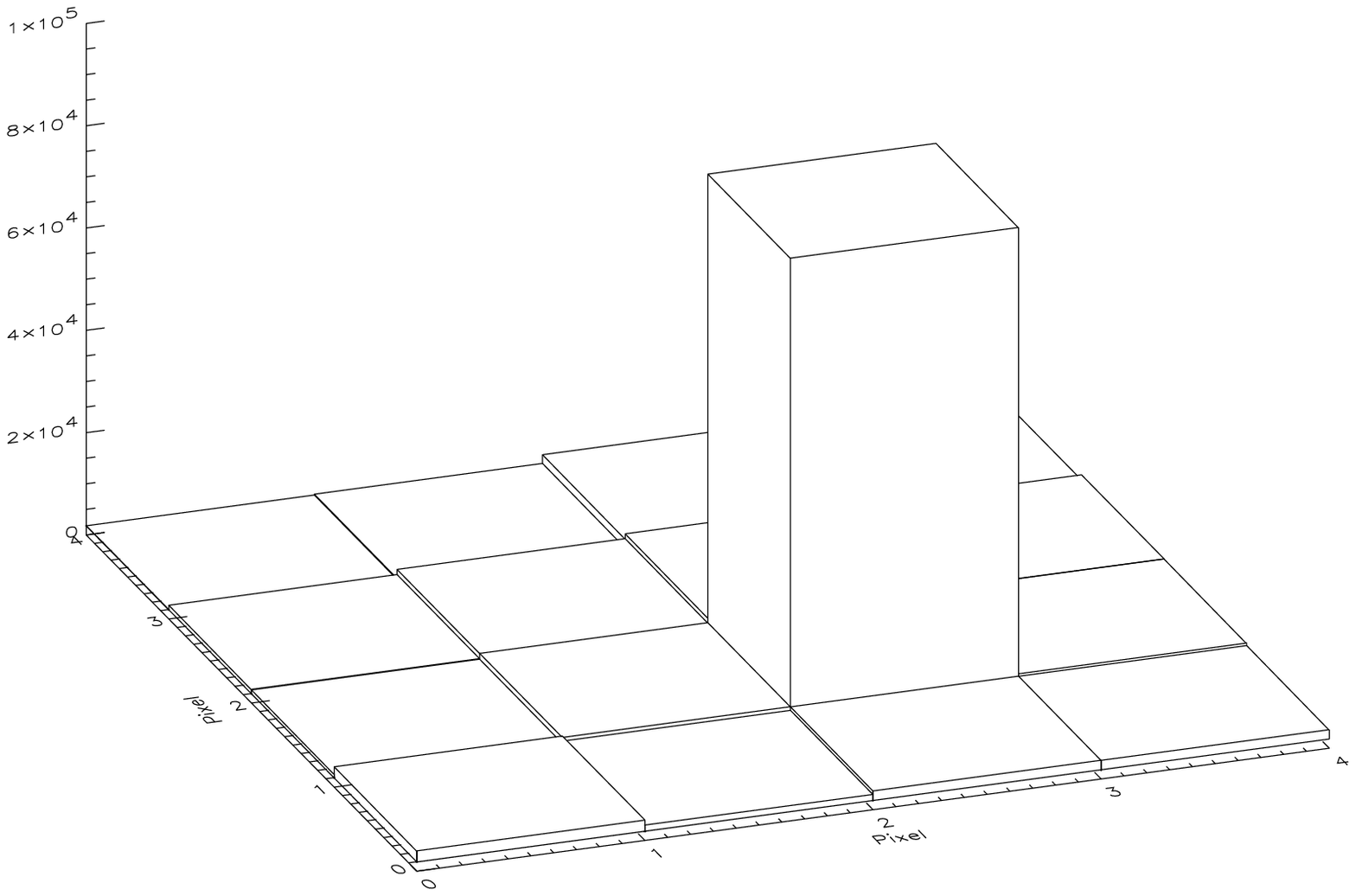,height=5cm} 
\psfig{figure=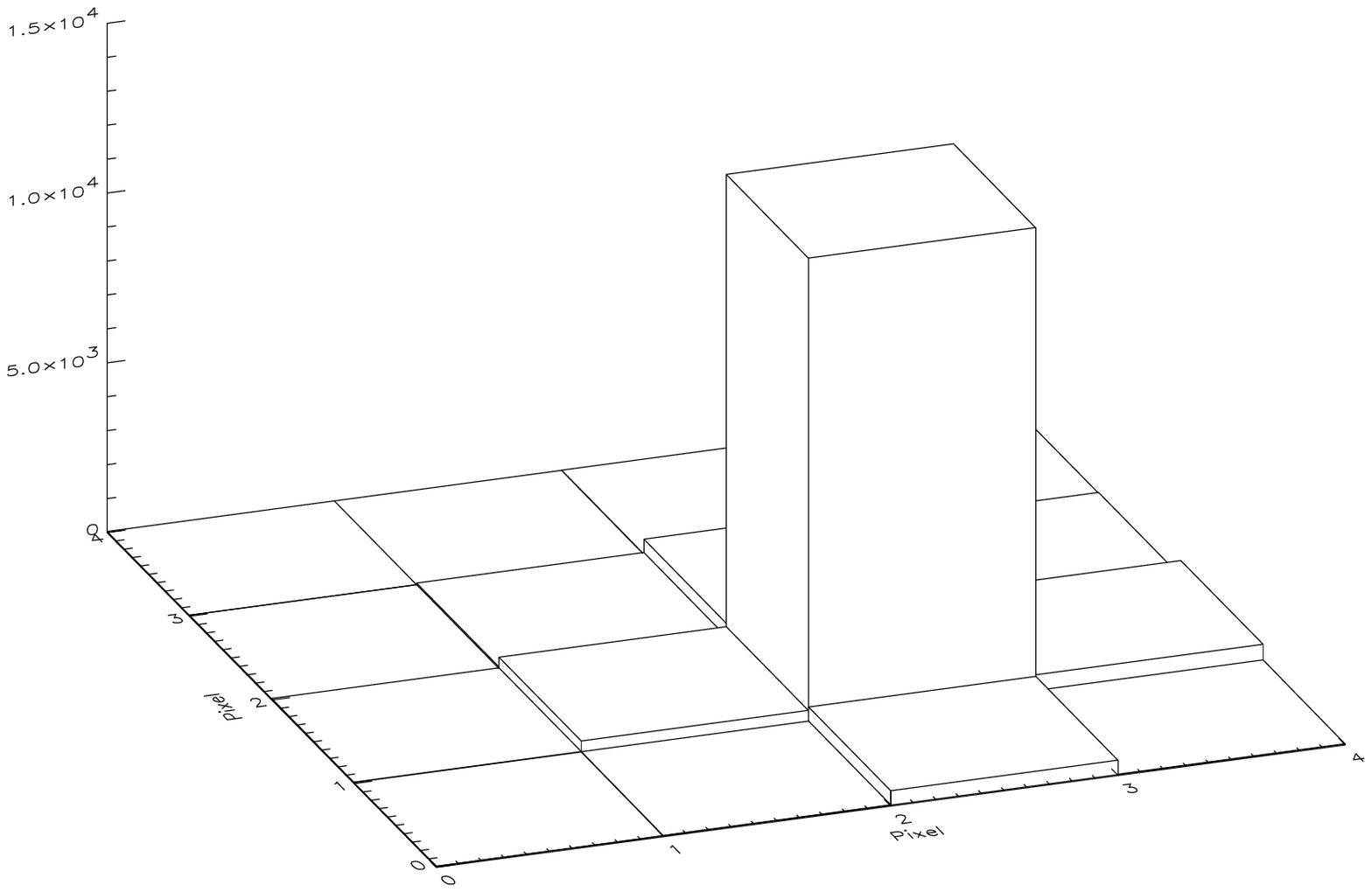,height=5cm} 
\end{tabular}
\end{center}
\caption[col_lego] 
{ \label{fig:col_lego}	  
The image of a collimated $^{57}$Co beam (left) and the extracted
image from one pixel under full-flood illumination (right). In the
collimated image, there is very little cross-talk between pixels.  The
extracted image shows the residual charge sharing among neighbor
pixels and the resulting PSF.}
\end{figure}

First, we investigated the amount of cross talk between pixels without
any contamination from photon interaction in the interpixel region.
Two spectra were recorded with $^{241}$Am and $^{57}$Co beams,
collimated to $\sim0.8$ mm diameter, centered on one pixel of the 5 mm
PIN detector.  In the $^{241}$Am spectra, there was no evidence of
cross talk between the center irradiated pixel and any of its nearest
neighbors.  But when we used a $^{57}$Co beam, we found a small
broadened photopeak at $\sim100$ keV energy in each of the four
nearest neighbor pixels at $\sim0.9$\% of the count rate of a center
pixel (Fig.~\ref{fig:col_pos}).  The far neighbor pixels (2 pixels from
center pixel) showed only the normal background and there was no sign
of a photopeak.  Figure~\ref{fig:col_lego} shows the total number of
counts in the center pixel compared to the neighbors for a collimated
beam.

We believe the photopeak cross talk is due to Compton scattered 122
keV and 136 keV photons.  The Compton scattering cross section at this
energy range is about one order of magnitude less than the
photoelectric absorption cross section.  The range of scattering angle
for a photon, which originally penetrates the CdZnTe at normal
incidence angle at the edge of the collimated beam and then scatters
into the next pixel, is $\sim30-100$ degrees.  This would correspond
to an observed near neighbor photopeak in the energy range of
$\sim115-90$ keV.  Although these interactions are unrecoverable, it
appears they will have a minimal effect on our PSF.

\begin{figure}
\begin{center}
\begin{tabular}{c}
\psfig{figure=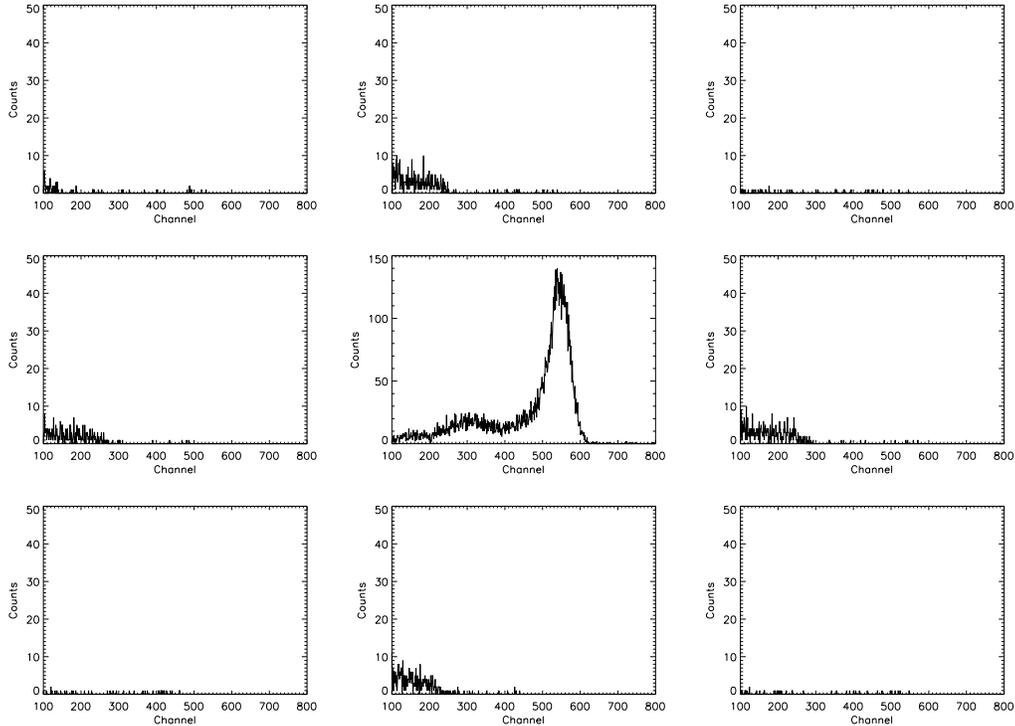,height=10cm} 
\end{tabular}
\end{center}
\caption[full_pos] 
{ \label{fig:full_pos}	  
The ``extracted'' (see text) spectrum from a full-flood $^{241}$Am
showing the effect of charge sharing between neighboring pixels.  The
low energy peak in channel $\sim300$ is the escape peak ($\sim30$
keV). Notice the lack of low energy tail below 25 keV in the center
pixel compared to the unextracted spectrum in
Figure~\ref{fig:energy}.}
\end{figure}

\begin{figure}
\begin{center}
\begin{tabular}{c}
\psfig{figure=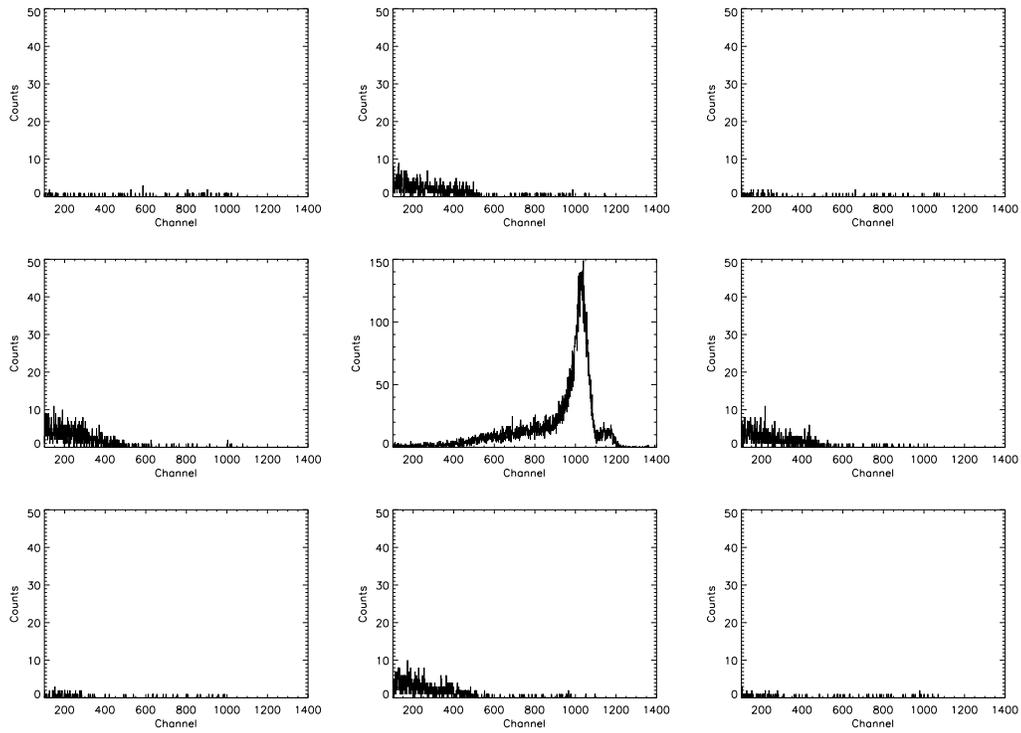,height=10cm} 
\end{tabular}
\end{center}
\caption[full_Co] 
{ \label{fig:full_Co}	  
The ``extracted'' (see text) spectrum for a full-flood $^{57}$Co.
The small Compton scattered photopeaks seen in the collimated spectrum
(Fig.~\ref{fig:col_pos}) do not appear in this extracted spectrum.
Instead, we are seeing the spread of the photoelectron generated charge cloud.
}
\end{figure}

Next, we measured the amount of cross talk due to charge sharing
between pixels from photon interaction in the interpixel region.  We
recorded simultaneous 16 channel spectra with full-flood beam of
$^{241}$Am on the 5 mm detector.  From the spectra, we extracted only
those events which had their highest pulse height on a given center
pixel.  An ``extracted'' spectrum should show the distribution of
events where the majority of the energy was deposited in this center
pixel and remaining charge, if any, spread among the neighbors.  The
resulting $^{241}$Am spectra (Fig.~\ref{fig:full_pos}) show each of the
near neighbors with a residual low energy tail and the center pixel
missing its portion of the low energy tail.  The missing low energy
counts are presumably the amount of charge normally shared from the
neighbor pixels.  The sum of the counts found in the near neighbor
spectra is $\sim20$\% of the counts in the center pixel photopeak.
The far neighbor pixels did not show any of sign of the residual tail.
The extracted full-flood $^{57}$Co spectra (Fig.~\ref{fig:full_Co})
show that $\sim36\%$ of the counts in the 122 keV photopeak is spread
among the near neighbors as low energy residuals.

Our simple interpretation is that we are observing charge cloud
sharing between pixels.  The effective pixel area extends into the
interpixel region where it can collect some fraction of the charge
cloud generated by an incident photon.  These low energy tails do not
appear in the collimated image (Fig.~\ref{fig:col_pos}) since the
collimated beam would not have produced charge clouds near the pixel
edge where it can be shared between pixels.  If the size of a charge
cloud from a 60 keV photon is on the order of our interpixel spacing
(0.2 mm), we estimate that $\sim25\%$ of the events occuring in a
pixel will have a fraction of its energy deposited in one of its near
neighbor pixels.  The lack of counts in the far neighbor pixels
indicate that our PSF is fairly tight (Fig.~\ref{fig:col_lego}).  Thus
any post-detection processing, such as summing signals of adjacent
pixels to recover the total deposited energy, can be done on our
detector with at most nine pixels.

%%-----------------------------------------------------------
\subsection{Effect of Guard Ring Bias}
\label{sect:guardbias}

\begin{figure}
\begin{center}
\begin{tabular}{c}
\psfig{figure=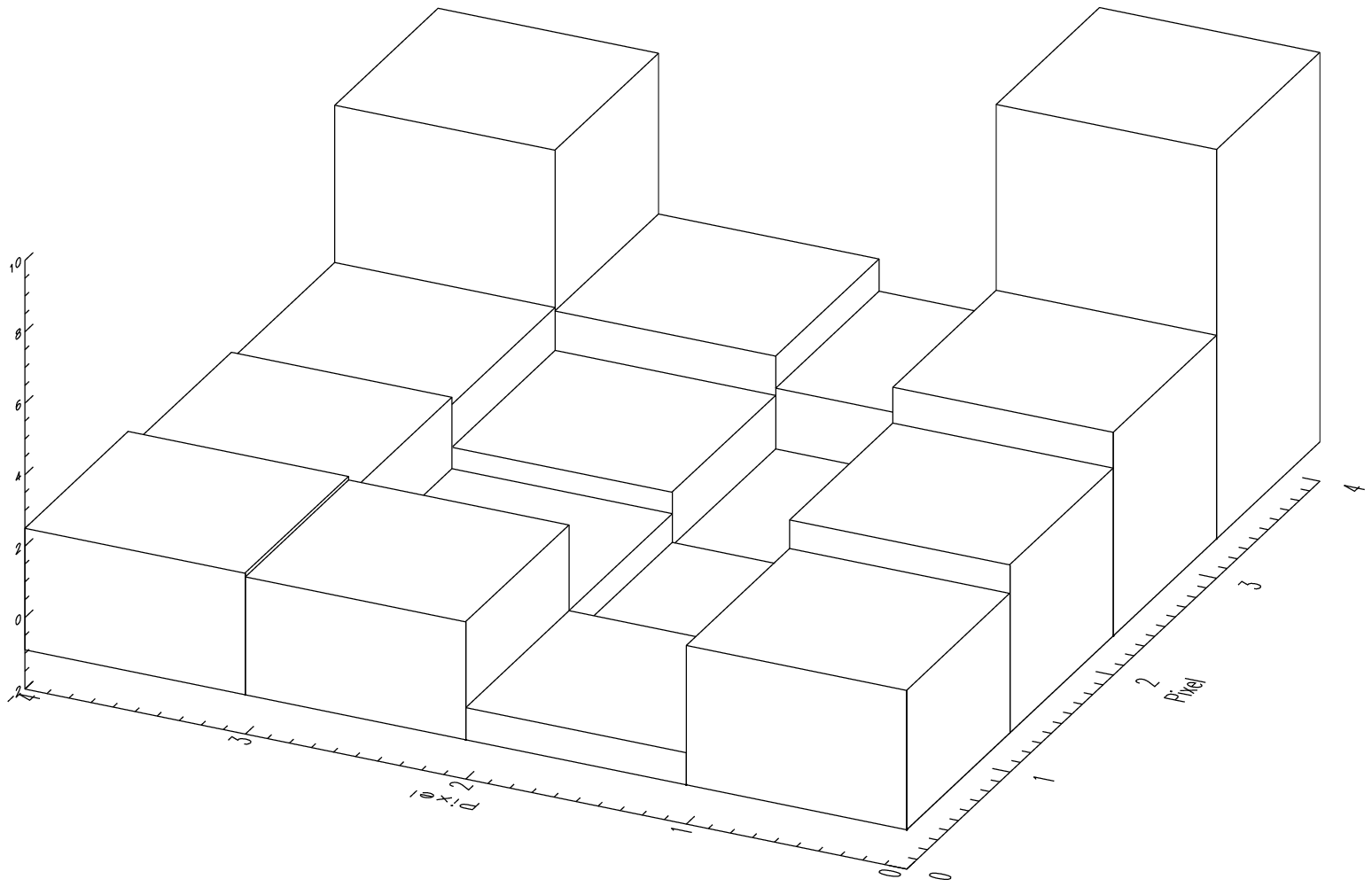,height=10cm} 
\end{tabular}
\end{center}
\caption[guard_bias] 
{ \label{fig:guard_bias}	  
Individual pixel improvements in photopeak efficiency when the outer guard
ring is biased with a steering potential.  The uncertainty in the
photopeak fit is $\pm0.5\%$ ($1\sigma$).  One of the foreground edge
pixel (6) was not functioning.}
\end{figure}

Another method of increasing photopeak efficiency is placing a
non-ground bias potential on the guard ring around the pixels.  The
additional potential sets up an electric field to deflect the
electrons toward the pixels, increasing their collection efficiencies.
This method has proved effective for single element
detectors\cite{butler97} and has been investigated for strip
detectors\cite{slavis98}.  To see whether a similar effect could be
seen on a pixellated detector, we full-flood illuminated the 5 mm PIN
detector with $^{241}$Am source and changed the guard ring potential
from ground to -25 volts with the detector cathode biased at -500
volts.  The guard ring surrounds the full $4\times4$ array and hence
it is only adjacent to the 12 pixels around the edge.  The
resulting spectra were analyzed for any change in efficiency or energy
resolution, especially for the edge pixels next to the guard ring.  We
found that the photopeak efficiencies of the inner four pixels, which
are surrounded by other pixels, averaged $-0.3\pm0.5\%$ change.  There
were no changes in their energy resolutions.  The edge pixels, which
are bound on one side by the guard ring, averaged $2.5\pm0.5\%$
improvement.  The four corner pixels, surrounded on two sides by the
guard ring, had their photopeak efficiencies improved on average by
$5.5\pm1.7\%$.  The complete detector response is shown in
Figure~\ref{fig:guard_bias}.

The general improvement in the photopeak efficiencies, especially for
the outer pixels, seem to indicate that additional guard ring bias
could be effective for pixellated detectors.  We did not use a bias
greater than 25 volts because two of the pixels (including one corner)
exhibited partial breakdowns across the narrow gap (0.2 mm) between
the pixel and the guard ring.  We are currently investigating the
cause of this effect.  As noted in Section~\ref{sect:leakage}, the
leakage current between a pixel and a guard ring is larger for the MSM
detector than for the PIN detector.  Therefore, we believe the use of PIN
blocking contacts will enhance the efficiency of a pixellated detector
with a biased guard ring.  For our current pixel mask geometry, the
guard ring is adjacent to at most two sides of the the outer pixels.
Such a design may be improved by placing a guard ring around each
pixel. We are currently testing a device which incorporates this new
design and will allow larger bias potentials for the guard ring.

%%%%%%%%%%%%%%%%%%%%%%%%%%%%%%%%%%%%%%%%%%%%%%%%%%%%%%%%%%%%%
\section{CONCLUSIONS} 
\label{sect:conclusion}  % \label{} allows reference to this section

The results presented in this paper show the potential for using PIN
contacts on CdZnTe for use in imaging hard X-ray detector.  The use of
blocking contacts show the anticipated improvement in the detector leakage
current compared to the standard MSM detector.  With the addition of
blocking contacts, it may be possible to economically construct a
large array detector using lower resistivity CdZnTe.  The two PIN
detectors we tested gave similar energy resolutions, and the pixel to
pixel responses were uniform across the detectors.  We did show
however, that the 2 mm and 5 mm thick detectors with same pixel size
gave very different photopeak efficiencies.  We attribute this to the
effectiveness of the small-pixel effect even for relatively large (1.5 mm)
pixels.  When determining the spatial resolution of our detector, we
found that the amount of charge spreading across pixels was relatively
small.  We will need to consider at most the nearest neighbor
pixels for any post-detection centroiding.  Finally, we found
that biasing the guard ring with a small steering potential is
effective for pixellated detectors.

Our goal is to eventually build a 1 m$^{2}$ pixellated detector array
for a hard X-ray imaging telescope.  We plan to construct this
detector using an array of interchangeable modular basic detector
elements (BDE), which consists of $2\times2$ array of 1 cm$^{2}$
CdZnTe\cite{grindlay98}.  Each BDE will be designed to operate with a
single 64 channel self-triggering ASIC and it will be made compact to
allow for seamless tiling.  We are presently testing a prototype BDE
which consists of two side by side 1 cm$^{2}$ CdZnTe detectors
epoxy bonded onto a ceramic substrate (flip-chipped), and read out by
a single 32 channel self-triggering ASIC (VA/TA from IDE).  Along with
the lab assessment of the detector, we also plan to study the CdZnTe
background in the near-space environement when we integrate the
prototype BDE with the next EXITE2 balloon flight, scheduled for
Spring 1999.

%%%%%%%%%%%%%%%%%%%%%%%%%%%%%%%%%%%%%%%%%%%%%%%%%%%%%%%%%%%%%
\acknowledgments     %>>>> equivalent to \section*{ACKNOWLEDGMENTS}       

We thank Einar Nyg\r{a}rd of Integrated Device Electronics for advice and
assistance with the ASIC. 
This work was supported in part by NASA grant NAG5-5103.
%%%%%%%%%%%%%%%%%%%%%%%%%%%%%%%%%%%%%%%%%%%%%%%%%%%%%%%%%%%%%
%%%%% References %%%%%

  \bibliography{paper}   %>>>> bibliography data in paper.bib
  \bibliographystyle{spiebib}   %>>>> makes bibtex use spiebib.bst
 
  \end{document}